\documentclass[11pt]{article}

\usepackage[margin=1in]{geometry}
\usepackage{times}
\usepackage{microtype}
\usepackage{graphicx}
\usepackage{booktabs}
\usepackage{tabularx}
\usepackage{array}
\usepackage{amsmath,amssymb}
\usepackage{enumitem}
\usepackage{listings}
\usepackage{xcolor}
\usepackage{tikz}
\usepackage[section]{placeins}
\usetikzlibrary{arrows.meta, positioning, fit, calc, shapes.multipart}
\usepackage[numbers,sort&compress]{natbib}
\usepackage[colorlinks=true,linkcolor=blue,citecolor=blue,urlcolor=blue]{hyperref}

\newcolumntype{Y}{>{\raggedright\arraybackslash}X}
\newcommand{\system}{CodeFirstReview}
\newcommand{\grv}{Generated Review View}
\newcommand{\reviewpkg}{Review Package}

\lstdefinestyle{yaml}{
  basicstyle=\ttfamily\small,
  columns=fullflexible,
  keepspaces=true,
  frame=single,
  breaklines=true,
  showstringspaces=false,
  literate={-}{-}1
}

\title{Review the Code, Not the Story:\\
A Vision and Protocol for Code-First Peer Review}
\author{Jienan Chen\\
National Key Laboratory of Science and Technology on Communications\\
University of Electronic Science and Technology of China, Chengdu 611731, China\\
\texttt{jesson.chen@outlook.com}}
\date{June 3, 2026}

\begin{document}
\maketitle

\begin{abstract}
Peer review in computational fields remains centered on author-written manuscripts, even though the decisive evidence for many claims resides in executable code, data, configurations, and experiment pipelines. This manuscript-first workflow gives authors substantial control over narrative framing while leaving reviewers with limited time to inspect implementation details, reproduce results, or detect unsupported claims. This vision and protocol paper proposes \emph{code-first peer review}: authors submit executable research artifacts and minimal claim manifests; a venue-controlled AI system builds the environment, executes experiments, audits code paths, maps claims to evidence, and generates a standardized \reviewpkg{} for human reviewers. The goal is not to replace reviewers or to give authors an automatic writing assistant. Instead, AI serves as review infrastructure that shifts the target of peer review from polished narratives to executable evidence. We formalize a claim-evidence contract, define the \grv{} and \reviewpkg{} abstractions, give a worked example, outline a system architecture, and analyze evaluation and governance challenges including AI bias, prompt injection, model instability, auditability, and author appeal.
\end{abstract}

\section{Introduction}

Computational research in computer science and communication systems is increasingly artifact-dependent. A networking paper may rely on protocol implementations, traffic generators, channel models, simulation configurations, and packet traces. A systems paper may depend on a benchmark harness, hardware profile, compiler flags, and failure-handling logic. A machine learning paper may depend on training code, data preprocessing, random seeds, hyperparameter schedules, and baseline implementations. Yet peer review still largely treats the author-written manuscript as the primary object of evaluation. Code, data, and execution environments are often submitted as supplementary artifacts, if they are submitted at all.

This workflow creates a structural asymmetry. Reviewers are asked to judge scientific claims through an author-controlled narrative, while the decisive evidence often resides in executable artifacts. Implementation choices, baseline tuning, random seed selection, preprocessing decisions, exception handling, failed runs, and hidden assumptions may be invisible in the manuscript. Even when code is available, reviewers rarely have enough time to reconstruct the environment, run all experiments, inspect all baseline implementations, and verify all reported claims. The result is not merely a reproducibility problem; it is a review-interface problem. Peer review is often forced to review the story before it can review the evidence.

Prior work has already moved toward reproducible computational publishing. Jupyter notebooks integrate prose, code, and results, reducing the risk that code and prose diverge \citep{kluyver2016jupyter}. Executable research compendia package paper, source code, computational environment, data, and interface definitions \citep{nuest2017erc}. ACM artifact badging recognizes artifacts that are available, evaluated, and tied to result validation \citep{acmArtifactBadging}. More recently, AI agents have been explored for scientific writing, autonomous discovery, code generation from papers, and computational reproducibility \citep{harper2024automated,lu2024aiscientist,yamada2025aiscientistv2,seo2025paper2code,starace2025paperbench,siegel2024corebench}. These developments make a stronger institutional redesign plausible: rather than asking authors to submit polished papers and optional artifacts, venues can ask authors to submit executable artifacts first, then generate standardized review views from those artifacts.

We propose \emph{code-first peer review}. In this paradigm, authors submit source code, data or data-access scripts, environment specifications, experiment scripts, configuration files, baseline implementations, and a minimal claim manifest. A venue-controlled AI system ingests the submission, builds the environment, runs experiments, audits code paths, extracts results, maps claims to evidence, and generates a standardized \reviewpkg{}. The package includes a manuscript-like \grv{}, a reproducibility report, a claim-evidence matrix, a code audit report, a baseline fairness report, a limitation report, and a provenance log. Reviewers read and evaluate this package, but the final decision remains a human editorial and peer-review decision.

The key distinction is control. We are not proposing that authors use AI to write better papers. We are proposing that venues control the generation of the review-facing manuscript view. Authors may correct misunderstandings by revising the artifact or the claim manifest, but they cannot directly polish or manipulate the generated narrative. Manuscript generation becomes an interface, not the goal. The venue asks: \emph{what does the executable evidence support?}

This paper is intentionally a vision and protocol paper rather than a deployed-system report. Its contribution is to make the review interface precise enough that future systems, venues, and benchmarks can test it. In particular, the paper makes four contributions.
\begin{enumerate}[leftmargin=1.5em]
    \item We define code-first peer review as an artifact-first, venue-controlled alternative to manuscript-first review for computational and communication-systems research.
    \item We formalize a claim-evidence contract that binds declared claims to scripts, configurations, metrics, baselines, logs, caveats, and reproducibility status.
    \item We introduce the \grv{} and \reviewpkg{} abstractions and illustrate them with a worked example that shows how a claim is accepted, scoped, or rejected by evidence.
    \item We propose a system architecture, evaluation agenda, and threat model covering faithfulness, reproducibility, defect detection, reviewer workload, metadata bias, prompt injection, model-version stability, and author appeal.
\end{enumerate}

\section{Motivation: From Manuscript-First to Artifact-First Review}

\subsection{The manuscript-first bottleneck}

The traditional submission workflow can be summarized as follows: authors perform experiments, write a manuscript, optionally release code, and reviewers evaluate the manuscript. This order matters. Because the manuscript is authored before review and is controlled by authors, the review process is exposed to narrative selection. A paper may emphasize favorable configurations, de-emphasize weak baselines, omit failed experiments, or describe implementation details at a level that is too abstract to verify. Strong writing can compensate for weak evidence; weak writing can obscure strong evidence. In fields where results depend on code, this is a poor alignment between evidence and evaluation.

Artifact evaluation addresses part of this problem, but it is usually additional to paper review. ACM badging, for example, recognizes artifacts that are documented, consistent, complete, exercisable, and supported by evidence of verification and validation \citep{acmArtifactBadging}. The limitation is scalability. A full artifact review requires environment reconstruction, execution, provenance capture, and often substantial manual debugging. Many conferences lack enough artifact reviewers, and artifact checks may occur after technical acceptance, not as the primary review object.

\subsection{Why AI changes the review interface}

AI agents are not yet reliable enough to replace reviewers. Published benchmark results support this caution. PaperBench evaluates agents on replicating 20 ICML 2024 Spotlight and Oral papers and reports that the best tested agent achieved a 21.0\% average replication score \citep{starace2025paperbench}. CORE-Bench evaluates computational reproducibility agents across 270 tasks from 90 papers and reports that the best agent achieved 21\% accuracy on the hardest task \citep{siegel2024corebench}. These results do not imply that AI can autonomously judge science. They imply that AI can be useful as infrastructure for evidence extraction, while human reviewers retain authority over scientific judgment.

The opportunity is therefore not full automation of peer review. The opportunity is to change what reviewers see. Instead of reading an author-controlled narrative first and code second, reviewers can read a venue-generated evidence view that is grounded in executed artifacts.

\subsection{Venue-controlled generation}

A crucial design principle is that the generated review view should be controlled by the venue, not by the author. If authors can freely edit the generated manuscript, the system collapses back into manuscript-first review with AI-assisted writing. Under code-first peer review, authors can modify the underlying artifact, improve documentation, revise the claim manifest, or submit structured rebuttals. Any change triggers re-execution, regeneration, and audit logging. The author cannot directly rewrite the review-facing manuscript to hide limitations or alter emphasis without changing the underlying evidence.

\section{Related Work}

\paragraph{Executable and reproducible publishing.} Jupyter notebooks and executable research compendia show that computational publications can integrate code, outputs, data, and prose \citep{kluyver2016jupyter,nuest2017erc}. These systems primarily improve reproducibility and readability for readers and authors. Code-first peer review differs by moving artifact execution and manuscript-like representation to the venue side before technical review.

\paragraph{Artifact evaluation.} Artifact evaluation and badging policies formalize the availability, functionality, reusability, and validation of computational artifacts \citep{acmArtifactBadging}. Code-first peer review can be seen as a venue-scale extension: artifact review is not a separate badge-oriented process, but the first stage of technical review.

\paragraph{AI-generated scientific writing and discovery.} Harper explores automated article generation from Python code \citep{harper2024automated}. The AI Scientist and AI Scientist-v2 pursue end-to-end autonomous scientific discovery, including idea generation, code execution, paper writing, and simulated or real peer-review submission \citep{lu2024aiscientist,yamada2025aiscientistv2}. Our proposal differs in both agency and governance: human authors still conduct the research, but the venue controls the conversion of artifacts into review views.

\paragraph{Paper-to-code and replication benchmarks.} Paper2Code and related work transform papers into code repositories, addressing the lack of implementation artifacts \citep{seo2025paper2code}. PaperBench and CORE-Bench evaluate the ability of agents to replicate or reproduce research \citep{starace2025paperbench,siegel2024corebench}. Our direction is complementary: instead of reconstructing code from papers, venues require code first and generate review packages from it.

\paragraph{LLM-as-a-judge and AI peer-review risks.} LLM-as-a-judge systems face reliability, bias, robustness, and standardization challenges \citep{gu2024llmjudge}. Empirical work has reported affiliation bias in LLM-mediated review decisions \citep{vonwedel2024affiliation,vasu2025reviewbias}. Hidden prompts in manuscripts have also been reported as a way to manipulate AI-assisted peer review \citep{lin2025hiddenprompts,gibney2025hidden}. These risks are central to our governance design: AI may help generate evidence, but it must itself be audited.

\paragraph{Scholarly infrastructure and open review.} Code-first review is also related to broader efforts to redesign scholarly infrastructure. Open review, preregistration, reproducibility checklists, and artifact badges all try to reduce information asymmetry between authors, reviewers, and readers \citep{rosshellauer2017openpeerreview,nosek2018preregistration,stodden2016enhancing}. Our proposal differs by making executable artifacts and claim-evidence links the first-class objects that determine the review-facing representation.

\begin{table}[!htbp]
\centering
\small
\caption{Representative related directions and how code-first peer review differs.}
\label{tab:related}
\begin{tabularx}{\linewidth}{p{0.22\linewidth}Y Y}
\toprule
Direction & Primary object & Difference from code-first peer review \\
\midrule
Executable notebooks / compendia & Readable and executable research documents & Improve author-reader reproducibility but generally preserve author control over narrative. \\
Artifact evaluation & Paper plus submitted artifact & Adds artifact checks, often as a separate process; code-first review makes artifacts the primary review object. \\
AI writing from code & Code-to-manuscript generation & Helps produce articles; code-first review treats generation as a venue-controlled review interface. \\
Autonomous AI scientist & AI-generated ideas, code, experiments, and papers & Automates research production; code-first review keeps human research agency and changes peer-review infrastructure. \\
Paper-to-code systems & Papers converted into code repositories & Reconstructs missing implementations; code-first review requires executable artifacts at submission time. \\
LLM-as-a-judge & Model-generated judgments and scores & Code-first review uses AI for evidence extraction and audit, not as sole final authority. \\
\bottomrule
\end{tabularx}
\end{table}

\section{Problem Formulation}

We model a computational submission as an executable package
\begin{equation}
S = (C, D, E, X, B, M),
\end{equation}
where $C$ is source code, $D$ is data or data-access logic, $E$ is an environment specification, $X$ is a set of experiment scripts and configurations, $B$ is a set of baseline implementations or baseline references, and $M$ is a minimal claim manifest.

The venue-controlled system transforms $S$ into a review package
\begin{equation}
R = (G, Q, V, A, F, L, P),
\end{equation}
where $G$ is the \grv{}, $Q$ is a reproducibility report, $V$ is a claim-evidence matrix, $A$ is a code audit report, $F$ is a baseline fairness report, $L$ is a limitation report, and $P$ is a provenance log.

\subsection{Minimal claim manifest}

Raw code is not enough. Code can reveal implementation and results, but it cannot fully specify the intended contribution or research question. Authors therefore submit a minimal structured manifest. The manifest is not a manuscript. It is a compact declaration of claims, evidence scripts, metrics, baselines, hardware requirements, and known limitations.

\begin{lstlisting}[style=yaml,caption={Illustrative minimal claim manifest.},label={lst:manifest}]
problem: "Low-latency scheduling for edge communication systems"

main_claims:
  - id: C1
    statement: "The proposed scheduler reduces p95 latency under bursty traffic."
    evidence_script: "experiments/run_latency.py"
    metric: "p95_latency"
    baselines: ["FIFO", "EDF", "RoundRobin"]
    expected_direction: "lower_is_better"

  - id: C2
    statement: "The method maintains throughput under packet loss."
    evidence_script: "experiments/run_loss_sweep.py"
    metric: "throughput"
    baselines: ["TCP-baseline", "QUIC-baseline"]

hardware:
  cpu: "16 cores recommended"
  gpu: "not required"
  network_emulator: "Mininet or ns-3"

limitations:
  - "Evaluated on synthetic traffic traces."
  - "No real hardware deployment."
\end{lstlisting}

The manifest gives the system enough intent to avoid hallucinating a contribution, while restricting authors from crafting a persuasive free-form narrative.

\subsection{Generated Review View}

A \grv{} is a venue-controlled, AI-generated, manuscript-like representation of an executable artifact. It is designed for review, not for authorial persuasion. It contains an abstract-like summary, method summary, experiment summary, main results, claim-evidence references, limitations, and reproducibility status. Each nontrivial empirical statement should link to code, configuration, command, log, table, or output hash.

\begin{table}[!htbp]
\centering
\small
\caption{Traditional manuscripts versus \grv{}s.}
\label{tab:grv_vs_paper}
\begin{tabularx}{\linewidth}{p{0.24\linewidth}Y Y}
\toprule
Dimension & Traditional manuscript & \grv{} \\
\midrule
Controller & Author & Venue submission system \\
Editable by author & Yes & No direct editing; artifact or manifest revision only \\
Primary basis & Author narrative & Executed code, results, manifest, logs \\
Purpose & Persuasion, communication, archival record & Standardized review interface \\
Evidence linkage & Often implicit & Explicit claim-to-code and claim-to-result links \\
Main risk & Selective reporting, narrative manipulation, prose-code mismatch & AI misinterpretation, model bias, execution failure \\
Auditability & Limited & Versioned logs, hashes, prompts, model versions, commands \\
\bottomrule
\end{tabularx}
\end{table}

\subsection{Design goals}

A code-first review system should satisfy five design goals.
\begin{enumerate}[leftmargin=1.5em]
    \item \textbf{Faithfulness:} generated review views must not introduce unsupported claims or hide material limitations.
    \item \textbf{Reproducibility:} execution results must be tied to commands, environments, random seeds, hardware profiles, and output hashes.
    \item \textbf{Standardization:} reviewers should see comparable evidence structures across submissions.
    \item \textbf{Governability:} the system must be auditable for bias, prompt injection, and model-version drift.
    \item \textbf{Human authority:} AI should assist evidence extraction and review preparation; humans retain final authority over novelty, significance, and decisions.
\end{enumerate}

\section{Worked Example: From Artifact to Review Package}

To make the protocol concrete, consider a hypothetical networking submission that proposes an edge scheduler for bursty traffic. In a manuscript-first workflow, reviewers would first read a claim such as ``our scheduler reduces tail latency while preserving throughput'' and then decide whether to inspect the code. In code-first review, that sentence is not the primary object. The primary object is a claim-evidence contract that connects each claim to executable evidence and explicitly records the scope under which the claim holds.

\subsection{Submitted artifact}

The author submits a repository, a data directory, an environment file, and a manifest. A reduced version of the package is shown below.

\begin{lstlisting}[style=yaml,caption={Reduced artifact package for the worked example.},label={lst:worked_package}]
submission/
  Dockerfile
  requirements.txt
  manifest.yaml
  src/scheduler.py
  experiments/run_latency.py
  experiments/run_loss_sweep.py
  configs/bursty.yaml
  configs/loss_sweep.yaml
  data/synthetic_traces.csv
  baselines/fifo.py
  baselines/edf.py
\end{lstlisting}

The manifest declares three claims. The first two are empirical claims with declared scripts and metrics. The third is a broader deployment claim that the authors would normally be tempted to include in a polished paper.

\begin{lstlisting}[style=yaml,caption={Reduced claim manifest for the worked example.},label={lst:worked_manifest}]
claims:
  - id: C1
    statement: "The scheduler reduces p95 latency under bursty traffic."
    evidence_script: "experiments/run_latency.py"
    metric: "p95_latency_ms"
    baselines: ["FIFO", "EDF"]
    expected_direction: "lower_is_better"

  - id: C2
    statement: "The scheduler preserves throughput under packet loss."
    evidence_script: "experiments/run_loss_sweep.py"
    metric: "throughput_mbps"
    baselines: ["FIFO", "EDF"]
    expected_direction: "not_lower_than_baseline"

  - id: C3
    statement: "The scheduler generalizes to real 5G deployments."
    evidence_script: null
    metric: null
    baselines: []
\end{lstlisting}

\subsection{Execution and claim status}

The venue system builds the environment, runs the declared scripts, and records commands, seeds, output hashes, and failures. In this example, the latency script runs successfully for five seeds; the loss-sweep script runs for three seeds but one baseline fails at the highest loss rate; no real deployment traces or hardware scripts are present.

\begin{table}[!htbp]
\centering
\footnotesize
\caption{Worked example of claim-evidence mapping.}
\label{tab:worked_claims}
\begin{tabularx}{\linewidth}{p{0.06\linewidth}p{0.24\linewidth}Y p{0.23\linewidth}p{0.12\linewidth}}
\toprule
ID & Claim & Executed evidence & Caveat & Status \\
\midrule
C1 & Reduces p95 latency under bursty traffic. & \texttt{run\_latency.py}; five seeds; \texttt{bursty.yaml}; regenerated CSV and plot hashes. & Synthetic traces only. & Scoped \\
C2 & Preserves throughput under packet loss. & \texttt{run\_loss\_sweep.py}; three seeds; loss rates 0--40\%; FIFO and EDF baselines. & EDF failed at 40\% loss; fewer than five seeds. & Partial \\
C3 & Generalizes to real 5G deployments. & No deployment script, field trace, hardware profile, or attestation found. & Claim exceeds submitted evidence. & Unsupported \\
\bottomrule
\end{tabularx}
\end{table}

\subsection{Generated review-view excerpt}

The \grv{} should not rewrite the authors' broadest claim into a persuasive form. It should preserve the evidential boundary. A suitable generated excerpt would say:

\begin{quote}
The artifact supports a scoped latency claim: under the provided synthetic bursty traces and five random seeds, the proposed scheduler reduces p95 latency relative to FIFO and EDF baselines. The throughput claim is only partially supported because one baseline run fails at the highest loss setting and the seed count is limited. The submitted artifact does not support a claim of generalization to real 5G deployments because no deployment traces, hardware experiment, or remote attestation evidence is included.
\end{quote}

This example illustrates the intended shift. The system is not deciding whether the scheduler is novel or important. It is converting an executable artifact into a conservative review interface that tells reviewers what the artifact supports, what it only partially supports, and what it does not support.

\section{The Code-First Peer Review Protocol}

\begin{figure}[!htbp]
\centering
\resizebox{\linewidth}{!}{%
\begin{tikzpicture}[
    node distance=1.15cm,
    box/.style={draw, rounded corners, align=center, minimum width=2.6cm, minimum height=0.85cm, font=\small},
    arrow/.style={-Latex, thick}
]

\node[box] (a1) {Author\\conducts research};
\node[box, right=of a1] (a2) {Author writes\\polished manuscript};
\node[box, right=of a2] (a3) {Reviewers read\\author narrative};
\node[box, right=of a3] (a4) {Optional code\\or artifact check};
\node[box, right=of a4] (a5) {Decision};
\draw[arrow] (a1) -- (a2);
\draw[arrow] (a2) -- (a3);
\draw[arrow] (a3) -- (a4);
\draw[arrow] (a4) -- (a5);
\node[font=\bfseries\small, left=0.3cm of a1] {Current};

\node[box, below=1.55cm of a1] (b1) {Author submits\\artifact + manifest};
\node[box, right=of b1] (b2) {Venue AI builds\\and executes};
\node[box, right=of b2] (b3) {Venue AI audits\\and maps claims};
\node[box, right=of b3] (b4) {Generated\\review package};
\node[box, right=of b4] (b5) {Human peer\\review + decision};
\draw[arrow] (b1) -- (b2);
\draw[arrow] (b2) -- (b3);
\draw[arrow] (b3) -- (b4);
\draw[arrow] (b4) -- (b5);
\node[font=\bfseries\small, left=0.3cm of b1] {Proposed};
\end{tikzpicture}}
\caption{The proposed protocol shifts peer review from author-controlled manuscript-first review to artifact-first review with venue-controlled generation of review packages.}
\label{fig:workflow}
\end{figure}
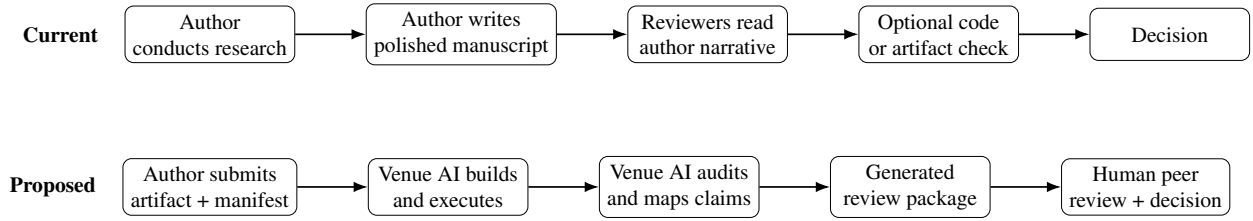

Figure~\ref{fig:workflow} summarizes the shift. The protocol has eight stages.

\paragraph{Stage 1: Submission ingestion.} The author uploads the executable package $S$. The system validates required files, license metadata, declared compute requirements, data access constraints, and manifest structure.

\paragraph{Stage 2: Environment reconstruction.} The system builds a container or equivalent environment from Dockerfiles, Conda files, Nix flakes, requirements files, continuous-integration scripts, README instructions, and manifest declarations. Failures are recorded rather than silently fixed. The system may attempt bounded dependency repair, but every repair becomes part of the provenance log.

\paragraph{Stage 3: Static and dynamic code audit.} The system constructs dependency graphs, identifies experiment entry points, traces metric computation, detects hardcoded outputs, compares baseline execution paths, and searches for data leakage or test-set tuning. Dynamic audit instruments execution to confirm which code paths generate reported results.

\paragraph{Stage 4: Experiment execution.} The system runs declared experiments under controlled budgets. It records commands, software versions, hardware profile, random seeds, resource usage, logs, generated files, and output hashes. For expensive experiments, the system may require a reduced verification suite plus a sampled full-scale audit.

\paragraph{Stage 5: Result extraction.} The system extracts metrics from logs, CSV files, databases, notebooks, plots, or structured outputs. It normalizes metrics and checks whether extracted results match the manifest's expected directions.

\paragraph{Stage 6: Claim-evidence mapping.} Each claim is mapped to code files, functions, experiment scripts, configurations, metrics, baselines, output artifacts, and caveats. Unsupported claims are marked as unsupported rather than paraphrased away.

\paragraph{Stage 7: Venue-controlled review-view generation.} The system generates $G$, the \grv{}, from the verified evidence. The generated text should be conservative: it may summarize results and methods, but it must not infer novelty, significance, or generality beyond the artifact and manifest.

\paragraph{Stage 8: Human review, rebuttal, and revision.} Reviewers evaluate the \reviewpkg{}. Authors may respond through structured rebuttals or artifact revisions. Revisions trigger regeneration and audit logging. The final accept/reject decision remains with human reviewers and editors.

\section{System Architecture}

\begin{figure}[!htbp]
\centering
\resizebox{\linewidth}{!}{%
\begin{tikzpicture}[
    node distance=0.9cm and 1.1cm,
    box/.style={draw, rounded corners, align=center, minimum width=3.0cm, minimum height=0.85cm, font=\small},
    wide/.style={draw, rounded corners, align=center, minimum width=3.8cm, minimum height=0.85cm, font=\small},
    arrow/.style={-Latex, thick}
]
\node[box] (input) {Executable\\Submission\\Package};
\node[box, right=of input] (repo) {Repository\\Analyzer};
\node[box, right=of repo] (env) {Build and\\Execution Agent};
\node[box, right=of env] (extract) {Result\\Extractor};
\node[box, below=of repo] (audit) {Code Audit\\Agent};
\node[box, below=of env] (claim) {Claim\\Verifier};
\node[box, below=of extract] (fair) {Fairness and\\Security Monitor};
\node[wide, right=1.2cm of extract] (package) {Generated\\Review Package};
\node[wide, right=1.2cm of package] (human) {Human Reviewer\\Interface};
\draw[arrow] (input) -- (repo);
\draw[arrow] (repo) -- (env);
\draw[arrow] (env) -- (extract);
\draw[arrow] (repo) -- (audit);
\draw[arrow] (env) -- (claim);
\draw[arrow] (extract) -- (claim);
\draw[arrow] (audit) -- (claim);
\draw[arrow] (claim) -- (package);
\draw[arrow] (extract) -- (package);
\draw[arrow] (fair) -- (package);
\draw[arrow] (package) -- (human);
\draw[arrow] (audit) -- (fair);
\draw[arrow] (claim) -- (fair);
\node[draw, dashed, rounded corners, fit=(repo)(env)(extract)(audit)(claim)(fair), inner sep=0.35cm, label={[font=\small\bfseries]above:Venue-controlled AI review infrastructure}] {};
\end{tikzpicture}}
\caption{A reference architecture for \system{}. The AI components are controlled by the venue and produce evidence for human review rather than final decisions.}
\label{fig:architecture}
\end{figure}
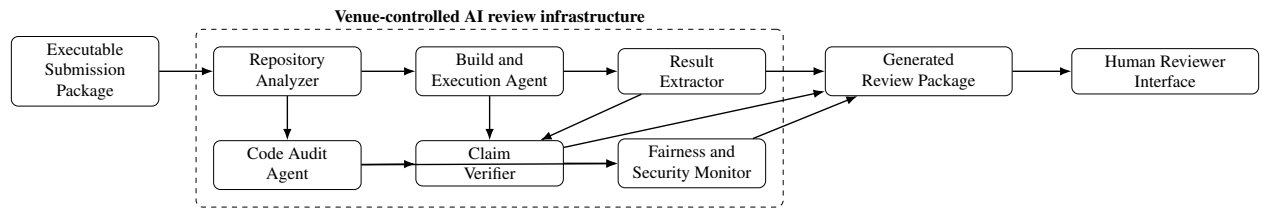

\subsection{Repository Analyzer}

The Repository Analyzer identifies source files, experiment entry points, configuration hierarchies, dependency specifications, scripts, notebooks, tests, and generated artifacts. It builds an experiment graph linking claims to scripts, scripts to configurations, configurations to data, and outputs to metrics. For communication-systems research, it should recognize common artifacts such as ns-3 simulations, Mininet topologies, OMNeT++ configurations, packet traces, channel-model scripts, and traffic-generation tools.

\subsection{Build and Execution Agent}

The Build and Execution Agent reconstructs the environment and runs experiments. It should operate in a sandbox, enforce resource quotas, and preserve exact execution traces. It should distinguish between author-declared dependencies, inferred dependencies, and system-added repairs. This distinction matters: an artifact that runs only after substantial AI repair should not receive the same reproducibility status as one that runs directly.

\subsection{Result Extraction Agent}

The Result Extraction Agent parses logs, tables, plots, notebooks, and output files. It verifies whether tables or figures can be regenerated from raw outputs. If plots are image-only, it should request or infer the underlying data only with uncertainty flags. For deterministic experiments, exact output hashes may be required; for stochastic experiments, confidence intervals, seed sweeps, and tolerance bands are more appropriate.

\subsection{Claim Verification Agent}

The Claim Verification Agent produces a claim-evidence matrix. An example row is shown in Table~\ref{tab:claim_matrix}. The matrix is central because it prevents the \grv{} from becoming an unsupported narrative.

\begin{table}[!htbp]
\centering
\footnotesize
\caption{Example structure of a claim-evidence matrix.}
\label{tab:claim_matrix}
\begin{tabularx}{\linewidth}{p{0.07\linewidth}p{0.27\linewidth}Y Y p{0.10\linewidth}}
\toprule
ID & Claim & Evidence & Caveat & Status \\
\midrule
C1 & Scheduler reduces p95 latency under bursty traffic. & \texttt{run\_latency.py}; config \texttt{bursty.yaml}; metric \texttt{p95\_latency}; output hash; baseline logs. & Synthetic traces only; no hardware deployment. & Scoped \\
C2 & Throughput remains stable under packet loss. & \texttt{run\_loss\_sweep.py}; loss rates 20--40\%; throughput CSV; seed sweep. & Fewer than five seeds; one baseline failed at 40\% loss. & Partial \\
C3 & Method generalizes to real-world 5G deployments. & No deployment scripts or traces found. & Claim exceeds artifact evidence. & Unsupported \\
\bottomrule
\end{tabularx}
\end{table}

\subsection{Fairness and Security Monitor}

The Fairness and Security Monitor checks for metadata sensitivity, prompt injection, suspicious instructions in comments or hidden text, model-version drift, and judge disagreement. It should blind author metadata by default. It should also treat repository text as untrusted input, because README files, comments, PDFs, notebooks, and metadata can contain instructions aimed at manipulating AI systems.

\subsection{Reviewer Interface}

Reviewers should not merely receive an AI-written paper. They should receive an interactive or static package that exposes evidence. A reviewer should be able to click from a sentence in the \grv{} to the corresponding claim, code path, command, log, metric, and generated file. The interface should support filters such as ``unsupported claims,'' ``failed experiments,'' ``baseline differences,'' and ``limitations inferred by the system.''

\section{Motivating Data from Prior Work}

Table~\ref{tab:motivation_data} collects data points from published work that motivate the protocol. These are not results from our proposed system; they are secondary evidence showing that (i) AI reproducibility agents remain imperfect, (ii) artifact standards already exist, and (iii) AI-assisted review introduces fairness and security risks that must be governed.

\begin{table}[!htbp]
\centering
\small
\caption{Public data points motivating code-first peer review.}
\label{tab:motivation_data}
\begin{tabularx}{\linewidth}{p{0.24\linewidth}Y Y}
\toprule
Source & Reported data point & Design implication \\
\midrule
PaperBench \citep{starace2025paperbench} & Best tested agent achieved a 21.0\% average replication score. & AI can assist evidence extraction, but should not replace reviewers. \\
CORE-Bench \citep{siegel2024corebench} & Best agent achieved 21\% accuracy on the hardest computational-reproducibility task. & Environment reconstruction and execution remain difficult review tasks. \\
JAMA affiliation-bias study \citep{vonwedel2024affiliation} & The study ran 232,500 LLM reviews and reported acceptance rates of 38.4\% for top-tier affiliation, 37.5\% for mid-tier, 36.7\% for low-tier, and 40.0\% without affiliation. & Metadata blinding and counterfactual audits should be part of AI-mediated review. \\
Hidden-prompt commentary \citep{lin2025hiddenprompts} & 18 arXiv manuscripts were reported to contain hidden prompts intended to manipulate AI-assisted review. & Submission portals need prompt-injection screening and input sanitization. \\
ACM artifact badging \citep{acmArtifactBadging} & Functional artifacts are expected to be documented, consistent, complete, and exercisable. & Existing artifact standards can seed review-package checklists. \\
\bottomrule
\end{tabularx}
\end{table}

\section{Evaluation Methodology}

A code-first review system should be evaluated as review infrastructure, not as a text generator. We propose six research questions.

\paragraph{RQ1: Faithfulness.} Does the \grv{} faithfully represent the submitted artifact? Metrics include hallucinated-claim rate, unsupported-generalization rate, missing-limitation rate, method-description completeness, citation relevance, and claim-evidence link correctness.

\paragraph{RQ2: Reproducibility.} Can the system build the environment and run the core experiments? Metrics include build success rate, experiment success rate, result reproduction error, time-to-first-run, dependency repair count, and provenance completeness.

\paragraph{RQ3: Defect detection.} Can the system detect material problems? Candidate defects include missing dependencies, hardcoded results, data leakage, unfair baseline budgets, cherry-picked random seeds, miscomputed metrics, invalid statistical tests, and missing data files. Synthetic defect injection can provide ground truth.

\paragraph{RQ4: Reviewer workload and quality.} Does the \reviewpkg{} reduce reviewer time while improving evidence quality? A human study can compare reviewers assigned to traditional manuscripts versus reviewers assigned to generated packages. Outcomes include review time, number of evidence-backed criticisms, detection of unsupported claims, reviewer confidence, and agreement with artifact experts.

\paragraph{RQ5: Fairness and robustness.} Does the system's output change under irrelevant metadata or input ordering? Counterfactual audits should perturb author names, affiliations, countries, writing style, file order, README wording, and model version. A simple metadata sensitivity score can be defined as
\begin{equation}
\Delta_{meta} = \max_{m_i,m_j} \left| s(S,m_i) - s(S,m_j) \right|,
\end{equation}
where $s(S,m)$ is a review score or risk score for the same artifact $S$ under metadata condition $m$. For generated text, analogous distances can be computed over claim statuses, limitation lists, and review recommendations.

\paragraph{RQ6: Human judgment boundary.} Which judgments should remain human? We expect code-first systems to help with buildability, result extraction, claim support, and auditability. We do not expect them to fully decide novelty, taste, long-term significance, real-world deployment value, or community priority.

\begin{table}[!htbp]
\centering
\small
\caption{Evaluation metrics for code-first peer review systems.}
\label{tab:metrics}
\begin{tabularx}{\linewidth}{p{0.22\linewidth}Y Y}
\toprule
Dimension & Example metrics & Required evidence \\
\midrule
Faithfulness & Hallucinated-claim rate; unsupported-generalization rate; missing-limitation rate & Human-labeled claim-evidence pairs \\
Reproducibility & Build success; run success; reproduction error; dependency repairs & Logs, containers, hashes, hardware metadata \\
Defect detection & Precision/recall on injected bugs; false accusation rate & Synthetic defects and expert labels \\
Reviewer utility & Time saved; evidence-backed criticisms; reviewer confidence & Controlled reviewer study \\
Fairness & Metadata sensitivity; model-family disagreement; order sensitivity & Counterfactual metadata and prompt variants \\
Security & Prompt-injection detection; sandbox escapes; suspicious hidden instructions & Red-team submissions and static scans \\
\bottomrule
\end{tabularx}
\end{table}

\subsection{Dataset construction}

A realistic benchmark can be built from accepted papers with public artifacts in machine learning, systems, networking, and communication venues. Each sample should contain the original paper, repository, data-access instructions, environment specification, and available artifact-review notes. For each sample, expert annotators construct a gold claim-evidence matrix and identify core experiments. Additional synthetic variants introduce controlled defects. For communication systems, useful defect classes include mismatched channel assumptions, unrealistic traffic distributions, baseline queues with different scheduling budgets, missing random seeds in simulations, and results that depend on undocumented hardware or network-emulator settings.

\subsection{Baselines}

Baselines should include: (i) traditional manuscript-only review, (ii) artifact evaluation without AI assistance, (iii) author-controlled AI writing assistance, and (iv) venue-controlled code-first packages. The comparison should not only measure accept/reject agreement. It should measure whether reviewers can identify what the code actually supports.

\section{Governance and Threat Model}

AI-mediated peer review creates new institutional risks. Table~\ref{tab:threats} summarizes key threats and controls.

\begin{table}[!htbp]
\centering
\small
\caption{Threat model and safeguards.}
\label{tab:threats}
\begin{tabularx}{\linewidth}{p{0.23\linewidth}Y Y}
\toprule
Threat & Example & Safeguard \\
\midrule
Narrative manipulation & Author edits generated text to hide limitations & Venue-controlled generation; no direct manuscript editing \\
Artifact gaming & Code detects review environment and changes behavior & Sandboxed execution; randomized tests; static and dynamic audit \\
Prompt injection & README, PDF, comments, or hidden text instruct AI to give positive reviews & Treat all submission text as untrusted; scan hidden text; isolate system prompts \\
Metadata bias & Scores change with author institution or name & Metadata blinding; counterfactual fairness audits \\
Model-version drift & Different model versions produce different claim statuses & Version logging; calibration suites; multi-model adjudication \\
Compute inequality & Expensive experiments cannot be fully rerun by venue & Declared budgets; reduced verification suites; sampled full audits; hardware-normalized reporting \\
False accusation & AI incorrectly flags a bug or unsupported claim & Evidence-first flags; human review; author appeal; expert audit \\
Over-standardization & Nontraditional contributions are penalized because they lack standard benchmarks & Reviewer override; explicit scope labels; separate significance review \\
\bottomrule
\end{tabularx}
\end{table}

\subsection{Metadata blinding and counterfactual audits}

The venue system should remove author names, affiliations, countries, acknowledgments, and self-identifying paths before AI review wherever possible. Because code repositories can leak identity through usernames, commit history, file paths, and dataset URLs, blinding must apply beyond PDF metadata. Counterfactual audits should then re-run selected evaluations with synthetic metadata to estimate sensitivity.

\subsection{Prompt-injection defense}

Prompt injection is not hypothetical in scholarly review contexts. Hidden prompts in manuscripts have been reported, including instructions designed to manipulate AI-assisted peer review \citep{lin2025hiddenprompts,gibney2025hidden}. A code-first system must assume that every submitted text fragment is adversarial input. The AI system should separate instructions from evidence, sanitize hidden text, strip invisible or tiny-font content from PDFs, scan comments and notebooks, and use tool-level policies that prevent submitted text from overriding venue prompts.

\subsection{Human appeal and final authority}

Authors need an appeal mechanism because AI systems can misread code, fail to build correct environments, or overstate defects. Appeals should be structured: the author points to the artifact, manifest, execution log, or configuration that corrects the system's interpretation. If accepted, the correction triggers regeneration and a versioned audit log. Human reviewers and editors retain final authority over scientific significance and decision-making.

\section{Discussion}

\subsection{What code can and cannot prove}

Code can ground empirical claims. It can show how a metric was computed, which baseline was used, whether a result can be regenerated, and which assumptions are embedded in configurations. Code cannot fully determine scientific meaning. It cannot decide by itself whether a problem is important, whether a contribution changes the direction of a field, whether a simplified model is acceptable, or whether a negative result is valuable. Code-first review should therefore be framed as evidence-grounded peer review, not automated scientific judgment.

\subsection{Applicability to communication systems}

Communication and networking research is a strong target for code-first review because many claims are simulation- or benchmark-based. The protocol can expose channel models, packet-loss assumptions, traffic distributions, scheduler implementations, congestion-control baselines, topology files, emulator settings, and hardware constraints. It is especially useful for papers whose claimed improvements depend on many interacting implementation choices.

At the same time, some communication-system experiments depend on hardware testbeds, licensed spectrum, proprietary traces, or private industrial deployments. For these submissions, the venue may require partial execution, remote attestation, confidential data enclaves, or independent third-party artifact execution. Code-first review should not require public disclosure of sensitive data; it should require auditable evidence under appropriate access controls.

\subsection{Transition path}

Venues need not adopt code-first review all at once. A practical transition has four levels.
\begin{enumerate}[leftmargin=1.5em]
    \item \textbf{Level 0:} traditional manuscript review with optional artifact links.
    \item \textbf{Level 1:} required artifact checklist and machine-readable claim manifest.
    \item \textbf{Level 2:} venue-generated reproducibility report before reviewer assignment.
    \item \textbf{Level 3:} venue-generated \reviewpkg{} becomes the primary review interface.
    \item \textbf{Level 4:} final publication includes both an author-facing article and an archived venue-generated evidence package.
\end{enumerate}

\subsection{Publication after acceptance}

The artifact-first review view need not be the final published paper. After acceptance, authors and venues can decide how to publish: (i) the \grv{} plus an author-edited communication article, (ii) a traditional article plus archived \reviewpkg{}, or (iii) an executable article where claims remain linked to evidence. The key is that technical acceptance should be based on venue-controlled evidence views, not solely on author-controlled manuscripts.

\section{Limitations}

This paper is a vision and protocol proposal, not a completed deployed system. Several open questions remain. First, current agents have limited reproducibility performance, so early systems should focus on evidence extraction and reviewer assistance. Second, large experiments may be too expensive to rerun fully. Third, artifact-first requirements may disadvantage theoretical work, qualitative work, or research whose evidence is not primarily computational. Fourth, venue-controlled generation could introduce new biases if models are not transparent and calibrated. Fifth, authors may learn to game the infrastructure, requiring continuous red-teaming and policy updates.

These limitations do not undermine code-first review; they define its engineering and governance agenda.

\section{Conclusion}

Computational peer review should review executable evidence before reviewing polished stories. We proposed code-first peer review, a venue-controlled protocol in which authors submit artifacts and minimal claim manifests, AI systems execute and audit those artifacts, and reviewers evaluate standardized evidence packages. Manuscript generation is not the goal; it is an interface. By shifting control from author-crafted narrative to venue-generated evidence views, code-first review can improve reproducibility, expose hidden implementation details, reduce reviewer burden, and make unsupported claims easier to detect. The same shift also creates new risks: AI bias, prompt injection, model drift, false accusations, and over-standardization. The appropriate response is not to avoid AI in review infrastructure, but to govern it: blind metadata, audit models, log provenance, preserve author appeal, and keep human reviewers responsible for scientific significance and final decisions.

\appendix
\section{Example Review Package Contents}

A minimal \reviewpkg{} could contain the following files.
\begin{lstlisting}[style=yaml]
review_package/
  generated_review_view.pdf
  reproducibility_report.json
  claim_evidence_matrix.csv
  code_audit_report.md
  baseline_fairness_report.md
  limitation_report.md
  provenance/
    commands.log
    environment.lock
    hardware.json
    output_hashes.txt
    model_versions.json
    prompt_templates.txt
  author_rebuttal_form.md
\end{lstlisting}

\section{Example Reviewer Checklist}

\begin{enumerate}[leftmargin=1.5em]
    \item Which main claims are fully supported, partially supported, or unsupported by executed evidence?
    \item Did the system reproduce the core tables and figures under declared resource budgets?
    \item Are the baselines implemented fairly and run under comparable conditions?
    \item Are the limitations material enough to weaken the contribution?
    \item Does the work remain novel and significant after unsupported claims are removed?
    \item Did the AI system produce any questionable interpretation that requires human correction?
\end{enumerate}

\section{Reproducibility Statement for This Draft}

The CSV file \texttt{data/motivation\_data.csv} records the secondary data values used in Table~\ref{tab:motivation_data}. These values are for motivation only and are not original experimental results from \system{}.

\section{AI Tool Use Statement}

AI-assisted editing and LaTeX checking tools were used during preparation of this draft. The author reviewed the manuscript and remains responsible for its claims, citations, wording, and any errors.

\bibliographystyle{plainnat}
\bibliography{references}

\end{document}